\documentclass[a4paper]{article}

\usepackage[T1]{fontenc}

\usepackage{amsmath,amsthm,amscd,amssymb}
\usepackage{verbatim}
\usepackage{authblk}
\usepackage{hyperref}

\newcommand{\R}{\mathbb{R}}

\newcommand{\Z}{\mathbb{Z}}

\title{Diddy: a Python toolbox for infinite discrete dynamical systems}

\author{Ville Salo\footnote{ORCID: 0000-0002-2059-194X} }
\author{Ilkka Törmä\footnote{ORCID: 0000-0001-5541-8517}}
\affil{Department of Mathematics and Statistics \\ University of Turku \\ Turku, Finland \\ \texttt{\{vosalo,iatorm\}@utu.fi}}

\begin{document}
\maketitle

\begin{abstract}
  We introduce Diddy, a collection of Python scripts for analyzing infinite discrete dynamical systems.
  The main focus is on generalized multidimensional shifts of finite type (SFTs).
  We show how Diddy can be used to easily define SFTs and cellular automata, and analyze their basic properties.
  We also showcase how to verify or rediscover some results from coding theory and cellular automata theory.

\end{abstract}

\section{Introduction}

This paper introduces and showcases \emph{Diddy}, a new Python library and domain-specific language (DSL) for defining and analyzing infinite discrete dynamical systems.
Its main purpose is to facilitate research of concrete multidimensional shifts of finite type (SFTs) and cellular automata (CA), but the authors' intent is to extend it to encompass e.g.\ classes of finite graphs, geometric tilings and substitution systems.
Diddy is free and open source, and available at \cite{diddyrepo}.

The Diddy project arose from the authors' previous research, which has lately been characterized by computer-assisted proofs of combinatorial and dynamical properties of discrete objects. \cite{SaTo22goe,SaTo22uniqf,SaTo23}
Such research would greatly benefit from a flexible language for defining multidimensional subshifts and cellular automata (instead of working directly with cumbersome lists of forbidden patterns or local rules), a unified interface to a SAT solver and other auxiliary programs, and the ability to easily switch between a special-purpose language and Python when needed.
Diddy can be used either as a standalone interpreted language or as a Python library; in this article, we focus on the former.

Diddy is a work in progress and under rapid development.
We do not promise that future versions will be compatible with the sample code in this document.
At the time of writing, the main features of Diddy are:
\begin{itemize}
\item
The ability to define SFTs and CAs with first-order logical formulae, and by composing other CAs.
\item
Tests for SFT containment and CA equality.
\item
Computation of upper and lower bounds for the topological entropy and the minimum asymptotic density of a configuration of an SFT.
\item
A visualizer for the patterns of an SFT, with the ability to automatically complete a small pattern into a larger one.
\end{itemize}

\section{Definitions}

Let $H$ be a group generated by a finite set $S \subset H$.
A graph $G = (V, E)$ is \emph{$H$-like}, if $H$ acts freely on $G$ by graph automorphisms and the nodes are divided into finitely many $H$-orbits.
This means that there is a finite set $R \subset V$ of \emph{representative} nodes, and the edge set $E$ is completely determined by the set of edges with at least one endpoint in $R$.

We concentrate on the case $G = \Z^d$ in what follows.
For $\vec n \in \Z^d$, the translate $\vec n \cdot R$ of $R$ is called a \emph{cell}, and the cells form a partition of $V$.
All Diddy objects live on top of a $\Z^d$-like graph with a fixed set of representatives, called the \emph{topology}.
For most of the definitions below, one can assume $V = \Z^d$ and $E = \emptyset$, which corresponds to the standard setting of symbolic dynamics used in e.g.\ \cite{LiMa95}.
General $\Z^d$-like graphs are convenient in modeling certain objects in Diddy.

Let $A$ be a finite set, called the \emph{alphabet}.
A \emph{configuration} on a graph $G = (V, E)$ is a function $x : V \to A$ that labels each node with an element of $A$.
If $G$ is a $\Z^d$-like graph, then $\Z^d$ acts on the set of all configurations $A^V$ by translation: $(\vec n \cdot x)_v = x_{\vec n \cdot v}$ for all $x \in A^V$, $v \in V$ and $\vec n \in \Z^d$.
We call $A^V$ the \emph{full $G$-shift}.
A configuration $x \in A^V$ is \emph{$\vec n$-periodic}, or \emph{periodic along $\vec n$}, if $\vec n \cdot x = x$.
It is \emph{totally periodic} if it is periodic along some vectors $\vec n_1, \ldots, \vec n_d \in \Z^d$ that span $\R^d$.

A \emph{finite pattern} over $A$ on $G$ is given by a finite \emph{domain} $D \subset V$ and a function $P : D \to A$.
We denote $D = D(P)$.
Every set $F$ of finite patterns on $G$ defines a \emph{$G$-subshift} $X_F \subset A^V$ as the set of configurations where no element of $F$ occurs:
\[
  X_F = \{ x \in A^V \mid \forall P \in F, \vec n \in \Z^d : (\vec n \cdot x)|_{D(P)} \neq P \}.
\]
If $F$ is finite, then $X_F$ is a \emph{shift of finite type}, or SFT.

Let $X \subset A^V$ and $Y \subset B^V$ be $G$-subshifts.
Let $R \subset V$ be the representative nodes.
A \emph{block map} is a function $f : X \to Y$ defined by a finite \emph{neighborhood} $N \subset V$ and a \emph{local rule} $F : A^N \to B^R$ by $f(x)_{\vec n \cdot r} = F((\vec n \cdot x)|_N)_r$ for all $r \in R$ and $\vec n \in \Z^d$.
A \emph{cellular automaton} is a block map from a full $G$-shift to itself.

Let $W : A \to \R$ be a function, which we interpret as giving \emph{weights} to the elements of $A$.
The \emph{upper density} of a configuration $x \in A^V$ is defined as
\[
  W(x) = \limsup_{k \to \infty} \frac{\sum_{\vec n \in [-k,k]^d} \sum_{r \in R} W(x_{\vec n \cdot r})}{|R|(2k+1)^d}.
\]
The \emph{minimum density} $W(X)$ of a subshift $X \subset A^V$ is $\inf_{x \in X} W(x)$.
It is known that the minimum density of a subshift $X$ can always be reached by a configuration $x \in X$ for which the limit superior in $W(x)$ is actually a limit.

Let $\mathcal{V}$ be a set of variables.
A \emph{Boolean formula} over $\mathcal{V}$ is either a variable $v \in \mathcal{V}$ or one of the forms $\neg \phi$, $\phi \vee \psi$, or $\phi \wedge \psi$, where $\phi$ and $\psi$ are Boolean formulas over $\mathcal{V}$.
A formula $\phi$ is \emph{satisfiable}, if there is an assignment $\pi : \mathcal{V} \to \{\mathrm{True}, \mathrm{False}\}$ such that $\phi$ evaluates to $\mathrm{True}$ when its variables are substituted with their $\pi$-values.
A formula is in \emph{conjunctive normal form} (CNF), if it has the form $\bigwedge_{i=1}^k \bigvee_{j=1}^{n_i} a_{i,j}$, where each $a_{(i,j)}$ is a variable or the negation of a variable.
Every formula is equivalent to a CNF formula.
The \emph{Boolean satisfiability problem} (SAT) is the following decision problem: given a Boolean formula in CNF, determine if it is satisfiable.
SAT is NP-complete, but modern SAT solvers are remarkably efficient at solving many kinds of real-world SAT instances.
Diddy uses the Glucose 4.1 solver \cite{glucose} through the PySAT library \cite{pysat}.

A \emph{(maximizing) linear program} is given by a set of \emph{variables} $\mathcal{V}$ and a set of \emph{constraints} of the form $a_1 v_1 + \cdots + a_n v_n \bowtie b$, where $\bowtie$ is one of $\leq$, $=$ or $\geq$, each $v_i$ is a variable, and $a_1, \ldots, a_n, b \in \R$ are constants.
To \emph{solve} the linear program for a variable $v \in \mathcal{V}$ means to find a valuation $\pi : \mathcal{V} \to \R$ for the variables such that each constraint holds and $\pi(v)$ is maximal.
There is an obvious variant that minimizes the value instead.
Efficient algorithms exist for solving linear programs.
Diddy uses the default solver of the Pulp library \cite{pulp}.

\section{Representations}

In Diddy, the topology $G = (V, E)$ is defined by fixing the finite set $R \subset V$ of representatives, and a finite set $E_0$ of triples $(r, s, \vec n) \in R^2 \times \Z^d$.
Then the nodes $V = \{ (\vec v, r) \mid \vec n \in \Z^d, r \in R\}$ and edges $E = \{ ((\vec m, r), (\vec m + \vec n, s)) \mid \vec m \in \Z^d, (r, s, \vec n) \in E_0 \}$ of $G$ are obtained by translating $R$ and $E_0$.
The action of $\Z^d$ is given by $\vec n \cdot (\vec m, r) = (\vec m + \vec n, r)$.
For example, the two-dimensional square grid can be specified by $R = \{a\}$ and $E_0 = \{(a, a, (0,1)), (a, a, (1,0))\}$.

In order to use SAT solvers to analyze symbolic dynamical objects, we must encode the relevant problems as SAT instances.
Diddy can represent an SFT $X \subset A^V$ over $G$ in two ways: as a concrete collection of forbidden patterns, or as a Boolean formula that represents the complement of such a collection.
First, an alphabet $A = \{a_0, \ldots, a_{m-1}\}$ of size $m$ is represented by a collection of $m-1$ Boolean variables $v_1, \ldots v_{m-1}$ together with the formula $\phi_A(v_1, \ldots, v_{m-1}) = \bigwedge_{1 \leq i < j < m} \neg v_i \vee \neg v_j$ stating that at most one of them can be true.
The interpretation is that the all-False assignment represents $a_0$, and the assignment of some $v_i$ as $\mathrm{True}$ represents $a_i$.

Suppose that $D \subset V$ is a finite set and $\psi$ is a Boolean formula over a variable set $\mathcal{V}$ that includes $v_{e, i}$ for $e \in D$ and $1 \leq i < m$ and possibly some auxiliary variables.
It defines an SFT $X_\psi \subset A^V$ as follows.
A configuration $x \in A^V$ is in $X$ if and only if for each $\vec n \in \Z^d$, there exists an assignment $\pi : \mathcal{V} \to \{\mathrm{True}, \mathrm{False}\}$ such that $\psi$ evaluates to $\mathrm{True}$, and for each $e \in D$, the assignment $\pi$ restricted to $v_{e, 1}, \ldots, v_{e, m-1}$ represents $(\vec n \cdot x)_e$.
In other words, we set the values of the variables $v_{e, i}$ according to the local pattern of $x$ at $\vec n$, and require that the partially evaluated $\psi$ is satisfiable.

The representation by a Boolean formula is used for every SFT, and the representation by forbidden patterns only when needed.
The reason is that converting a list of forbidden patterns $F$ into a CNF formula $\phi_F$ is straightforward, and the size (in computer memory) of $\phi_F$ is linear in that of $F$.
However, a Boolean formula may require an exponentially larger set of forbidden patterns.

To represent a block map, we need to specify the neighborhood $N \subset V$ and the local rule $F : A^N \to B^R$. It suffices to describe for each node $r \in R$ and symbol $b \in B$ the set of patterns $P \in A^N$ satisyfing $F(P)_r = b$. This can be coded as a Boolean formula analogously to what we did with SFTs. Thus a block maps is defined by $|R| |B|$ Boolean formulas.

\section{Defining topologies, SFTs and block maps}
\label{sec:defining}

Recall that a $\Z^d$-like topology $G = (V, E)$ is defined by specifying the set $R \subset V$ of representatives and a set $E_0$ of triples $(r, s, \vec n) \in R^2 \times \Z^d$.
In Diddy, the triples are named, and can be referred to by these names.
For example, the following code defines the two-dimensional hexagonal grid (which is also available via the built-in command \texttt{\%topology hex}).
The representatives are $R = \{0,1\}$, and there are six triples, named \texttt{lt}, \texttt{rt}, \texttt{up} and \texttt{dn}.
Two triples may have equal names, if they originate from distinct vertices.
We also illustrate defining alphabets, in this case the binary alphabet $A = \{0,1\}$.
\begin{verbatim}
%dim 2
%nodes 0 1
%topology
lt (0,0,0) (-1,0,1); lt (0,0,1) (0,0,0); rt (0,0,0) (0,0,1);
rt (0,0,1) (1,0,0); up (0,0,0) (0,1,1); dn (0,0,0) (0,-1,1)
%alphabet 0 1
\end{verbatim}

In Diddy, SFTs can be defined in two ways: by a list of forbidden patterns, or by a \emph{first order formula} (FO formula) that is compiled into an intermediate circuit representation and then a SAT instance.
The language of valid FO formulae, at the time of writing, consists of the following elements (which we will not define formally in this paper):
\begin{itemize}
\item
  Variables ranging over nodes of $V$, cells of $\Z^d$, symbols of $A$, or truth values.
\item
  Moving along edges:
  If \texttt{x} is a node variable whose value is of the form $n = (\vec m, r)$, and \texttt{m} is the name of a triple $(r,s,\vec n)$, then \texttt{x.m} is the node $(\vec m + \vec n, s)$.
  If \texttt{x} is a cell or node and \texttt{s} is the name of a node $s$, then \texttt{x.s} instead denotes the node $(\vec m, s)$ in the cell (of) $x$.
\item
  Equality and proximality:
  If \texttt{x} and \texttt{y} are node variables, \texttt{x @ y} means they are the same node, and \texttt{x \textasciitilde{} y} means they are adjacent in $G$.
  If \texttt{x} is a node variable and \texttt{y} is a symbol variable (or a literal symbol), \texttt{x = y} means that \texttt{x} has the symbol \texttt{y}.
  If \texttt{y} is a node variable, it means that the nodes have the same symbol.
  These can be negated as \texttt{x !@ y}, \texttt{x !\textasciitilde{} y} and \texttt{x != y}.
\item
  Logical connectives:
  Diddy has the prefix operator \texttt{!} (negation), and infix operators \texttt{\&} (conjunction), \texttt{|} (disjunction), \texttt{->} (implication) and \texttt{<->} (equivalence) with the usual semantics.
\item
  Restricted quantification:
  \texttt{Ey[x2]} defines a new node variable \texttt{y} that is existentially quantified over $B_2(x)$, the ball of radius 2 centered on the existing node variable \texttt{x} with respect to the path distance of $G$.
  There can be more than one restriction inside the brackets: for example, in \texttt{Ez[x2y1]} the variable \texttt{z} ranges over the union $B_2(x) \cup B_1(y)$.
  \texttt{Ay[x2]} is the analogous universal quantifier.
  \texttt{EC} and \texttt{AC} quantify cells instead of nodes.
\item
  Local definitions:
  \texttt{let func a b := a @ b | (a = 0 \& b = 0) in} defines an auxiliary two-argument formula \texttt{func}, checking that its two arguments are either the same node or both contain the symbol \texttt{0}.
  In the code that follows the definition, it can be invoked as \texttt{func x y}.
  Auxiliary formulas can have any number of arguments, which can be node or symbol variables.
\end{itemize}

An SFT can be defined with the command \texttt{\%SFT name formula}.
We illustrate the syntax with an example.
A \emph{radius-$r$ identifying code} \cite{KaChLe98} on a graph $G = (V, E)$ is a subset $C \subset V$ such that
\begin{itemize}
\item all $v \in V$ satisfy $B_r(v) \cap C \neq \emptyset$, and
\item all $v \neq w \in V$ satisfy $B_r(v) \cap C \neq B_r(w) \cap C$.
\end{itemize}
The idea is that the nodes in $C$ are ``sensors'' that send an alert if some event occurs at any node within distance $r$.
In case an event occurs, we would like to have at least one alert, and to be able to infer its position from the set of alerting sensors.
The SFT of radius-$1$ identifying codes 
can be defined as follows:
\begin{verbatim}
%SFT idcode Ao let cnbr u v := v=1 & u~v in
  (Ed[o1] cnbr o d) &
  (Ap[o2] p!@o -> Eq[o1p1] (cnbr o q & p!~q) | (cnbr p q & o!~q))
\end{verbatim}
We first define an auxiliary formula \texttt{cnbr u v}, which is true if \texttt{v} is in $C$ (modeled as having the symbol 1) and adjacent to \texttt{u}.
On the next line, we check that the universally quantified \texttt{o}, representing an arbitrary node, has a neighbor \texttt{d} in $C$.
Finally, we check a condition on all nodes \texttt{p} within distance 2 of \texttt{o}: if \texttt{p} and \texttt{o} are distinct, one of them should have a neighbor \texttt{q} in $C$ which is not a neighbor of the other.
Note that this definition is independent of the topology $G$.

Currently, Diddy has no syntactic support for general block maps. The syntax for defining a CA on a full shift is \texttt{\%CA name preimages} where \texttt{preimages} is a list of commands of the form \texttt{node symbol formula}. The formula describes when the local rule $F : A^N \to A^R$ of the CA should write the symbol \texttt{symbol} in the node \texttt{node}, which is an element of $R$.
Similarly as when describing SFTs, $N$ is simply the set of nodes that the Boolean formula references.

We again omit a more precise description, and illustrate this with an example instead.
Consider three cellular automata $L, R, F$ defined on a two-track binary alphabet, i.e.\ $A = \{0,1\}^2$, where $L$ and $R$ shift the top track to the left and right respectively, and $F$ adds the top track to the bottom track modulo 2.
All three are reversible, and their compositions form a group that is isomorphic to the \emph{lamplighter group} $\mathcal{L}$.
It is the wreath product $\Z_2 \wr \Z$, i.e.\ semidirect product of the groups $\bigoplus_{n \in \Z} \Z_2 \rtimes \Z$ where $\Z$ acts on the infinite direct product $\bigoplus_{n \in \Z} \Z_2$ by shifting.
An element of the lamplighter group can be thought of as an instruction for shifting a bi-infinite tape of bits and flipping finitely many tape cells.



\begin{verbatim}
%alphabet 0 1
%nodes top bot -- two tracks, top and bottom
%dim 1
%topology
rt (0, top) (1, top); rt (0, bot) (1, bot);
lt (0, top) (-1, top); lt (0, bot) (-1, bot)
%CA R -- partial right shift on the top track
top 1 ACo o.top.lt=1
bot 1 ACo o.bot=1
%CA L -- partial left shift on the top track
top 1 ACo o.top.rt=1
bot 1 ACo o.bot=1
%CA F -- add top track to bottom track
top 1 ACo o.top=1
bot 1 ACo (o.bot=1 | o.top=1) & (o.bot=0 | o.top=0)
\end{verbatim}

\section{Comparing SFTs}

Given two SFTs $X, Y \subset A^V$ on $G$, do we have $X = Y$?
This problem is in general undecidable for $d \geq 2$, even when $Y = \emptyset$ is fixed, as shown by Berger in \cite{Be66}.
Thus we have no hope of implementing total algorithms for testing equality of SFTs.
However, the following is true:
\begin{enumerate}
\item
If $X \subseteq Y$, then it can be verified computationally.
This is due to a compactness argument.
Let $F$ and $F'$ be sets of forbidden patterns for $X$ and $Y$.
If $X \subseteq Y$, there must exist a finite set $N \subset \Z^d$ such that if $x \in A^V$ satisfies $(\vec n \cdot x)|_{D(P')} \neq P'$ for all $P' \in F'$ and $\vec n \in N$, then $x|_{D(P)} \neq P$ for all $P \in F$.
\item
If there exists a totally periodic configuration $x \in X \setminus Y$, then it can be found computationally, simply by enumerating totally periodic configurations and checking whether they contain forbidden patterns of $X$ and/or $Y$.
In particular, $X \subseteq Y$ is decidable if $X$ has dense totally periodic points (as many natural examples do).
\end{enumerate}
This gives rise to a partial algorithm for checking the containment $X \subseteq Y$.
For increasing $k = 1, 2, \ldots$, check whether condition 1 holds with $N = [-k, k]^d$, returning ``yes'' if it does; otherwise check condition 2 for the periods $\vec n_i = (0, \cdots, 0, k, 0, \cdots, 0)$, returning ``no'' if such a periodic configuration is found.
This is essentially Wang's semi-algorithm for checking whether a set of colored square tiles can tile the infinite plane \cite{Wa61}.
The command \texttt{\%equal} performs the checks with a SAT solver, using the representations of SFTs by formulas.
Of course, the algorithm never terminates if $X \setminus Y$ is nonempty but contains no totally periodic configurations.

\section{Computing and comparing block maps}

Given two $G$-SFTs $X \subseteq A^V$ and $Y \subseteq B^V$ and block maps $f, g : X \to Y$, it is undecidable whether $f = g$ for the same reason as equality checking between SFTs: if the local rules of $f$ and $g$ always give a different output, then $f = g$ if and only if $X = \emptyset$.
However, if $X = A^V$ is a full shift, the problem becomes ``merely'' co-NP-complete, as we now have to check that there does not exist an input that one of the circuits of $f$ evaluates to $\mathrm{True}$ and that of $g$ to $\mathrm{False}$, or vice versa.
Diddy provides this functionality with the \texttt{\%equal\_CA} command.

In addition, one can compose cellular automata, either explicitly using the command \texttt{\%compose\_CA name CA\_list}, or by enumerating all compositions of a given set of CAs up to a given length and reporting which of them are equal, using \texttt{\%calculate\_CA\_ball bound filename CA\_list}.
The latter command writes its results into a log file.
Its name refers to the fact that it essentially analyzes the shape and size of a ball in the Cayley graph of the semigroup generated by the given automata.

Continuing the lamplighter example, consider the cellular automata $\alpha = F R^3 F L^5 F R^2$ and $\beta = L^2 F R^5 F L^3 F$.
In both compositions, the top track is added to the bottom track shifted by $-2$, $0$ and $3$ steps (in some order).
Hence, they should be equal.
We can check this relation in Diddy:
\begin{verbatim}
%compose_CA alpha F R R R F L L L L L F R R
%compose_CA beta  L L F R R R R R F L L L F
%equal_CA alpha beta
\end{verbatim}

Diddy also allows us to compute the order of cellular automata, at least if it is small. Consider the cellular automaton $f$ on $\Z$ with alphabet $\{0,1\}$ and neighborhood $\{0,1,2,3,4,5,6\}$ which maps
\begin{align*}
f(x)_0 = 1 \iff x_{[0,6]} \in \{&0111000, 1000100, 0111100, 1010110, 1111110, 0010001, \\
&0101001, 1101001, 0000101, 0010101, 0011101, 0000011,\\
& 1110011, 0011011, 1011011, 0111011, 1111011, 1000111, \\
& 1100111, 0010111, 1110111, 0001111, 0111111\}. \end{align*}
The first-named author suggested in an invited talk of the AUTOMATA 2017 conference that this CA might be nilpotent, and showed that its nildegree (first power $n$ such that $f^n$ maps everything to $0$) is at least $7$. The author has later showed with an ad hoc proof that the nildegree is at most $9$.

With Diddy we can compute the nildigree directly. The cellular automaton can be translated quite directly into Diddy code (with some lines omitted):
\begin{verbatim}
%dim 1
%alphabet 0 1
%nodes 0
%topology rt (0,0) (1,0); lt (0,0) (-1,0)
%CA f
0 1 Ao let x a b c d e f g :=
  o=a & o.rt=b & o.rt.rt=c & o.rt.rt.rt=d &
  o.rt.rt.rt.rt=e & o.rt.rt.rt.rt.rt=f & o.rt.rt.rt.rt.rt.rt=g in
x 0 1 1 1 0 0 0 | x 1 0 0 0 1 0 0 | [...] | x 0 1 1 1 1 1 1
\end{verbatim}
Here we define a predicate \texttt{x} that checks whether the origin has a particular word in its neighborhood, and then take the disjunction over the set from above.
Now we define also the zero CA and calculate the ball that this pair generates:
\begin{verbatim}
%CA zero 0 1 0=1
%calculate_CA_ball 10 outfile f zero
\end{verbatim}
The output file contains several lines, but the most relevant one is \texttt{zero = f f f f f f f}.
It states that the seventh power of $f$ is the zero CA. One can also calculate this much quicker with \texttt{\%compose\_CA seventh f f f f f f f} and \texttt{\%equal\_CA seventh zero}, if we already know (or correctly guess) the order.

\section{Minimum density}

In many applications, it is important to determine the minimum density $W(X)$ of an SFT $X \subset A^V$ with respect to some alphabet weights $W : A \to \R$.
Diddy includes two algorithms for approximating $W(X)$, one for finding upper bounds and another one for lower bounds.
Neither of them is guaranteed to find the true value, but in practice they can give good results.

\subsection{Upper bounds by periodic configurations}

We first describe the algorithm for finding upper bounds.
Choose a set of $d-1$ vectors $K = \{\vec n_2, \vec n_3, \ldots, \vec n_d\} \subset \Z^d$ such that $\{\vec e_1, \vec n_2, \vec n_3, \ldots, \vec n_d\}$ is a basis of $\R^d$, where $\vec e_1 = (1,0, \ldots, 0)$ is the first standard basis vector.
Let $X_K \subset X$ be the set of configurations that are periodic along $n_i$ for each $2 \leq i \leq d$.
Then $X_K$ is also an SFT and $W(X_K) \geq W(X)$.
Moreover, we claim that $W(X_K)$ is computable in a reasonably effective and parallelizable way.
In fact, we have already announced a special case in \cite{SaTo23}.
To compute upper bounds for $W(X)$ one just needs to choose suitable sets of periods $K$ and apply the algorithm.
Note that if $X$ happens to be aperiodic, then this algorithm will never give a result for any choice of $K$, and in any case its performance depends on how well the density $W(X)$ is approximated by periodic configurations.

In this algorithm we use the representation of $X$ by a set $F$  of forbidden patterns.
Let $D \subset \Z^d$ be a fundamental domain of the vectors $K \cup \{\vec e_1\}$, and denote $U = \{a_2 \vec n_2 + \cdots + a_d \vec n_d \mid a_2, \ldots, a_d \in \Z\}$.
Let $B = \bigcup_{\vec u \in U} \vec u + D$ be the \emph{border}, and call $B^+ = \{ (\vec b + p \vec e_1) \cdot r \mid p \geq 0, r \in R \}$ the \emph{right side} and $B^- = \{ (\vec b + p \vec e_1) \cdot r \mid p \leq 0, r \in R \}$ the \emph{left side} of the vertex set $V$, where $R \subset V$ is the set of representatives.
The set $X_K$ is essentially a one-dimensional SFT whose alphabet is $A^{D \cdot R}$, and the dynamics is translation by $\vec e_1$.
We represent it as a labeled digraph $G' = (V', E')$ as follows.
A vertex $v \in V'$ is a set of translated finite patterns $Q = \vec m \cdot P$ where $P \in F$ and $\vec m \in \Z^d$ that is invariant under translation by $U$: if $Q \in v$ and $\vec n \in U$, then $\vec n \cdot Q \in v$.
We also require that for each $Q \in v$, the domain $D(Q)$ intersects both $B^+$ and $B^-$.
Note that while each $v \in V'$ is technically an infinite set of finite sets, the $U$-invariance and the intersection property ensures that it can be encoded by a bounded amount of data.
In particular, $V'$ is a finite set.

Denote by $v_0$ the set of patterns $Q = \vec m \cdot P$ for $P \in F$ and $\vec m \in \Z^d$ with $D(Q) \subset B^+$ and $D(Q) \cap B^- \neq \emptyset$.
For $v_1, v_2 \in V'$ and $S \in A^{D \cdot R}$, there is an edge from $v_1$ to $v_2$ with label $S$ if and only if
\[
  v_2 = \{ -\vec e_1 \cdot Q \mid Q \in v_1 \cup v_0, \nexists n \in D(Q) \cap D(S) : Q_n \neq S_n \}.
\]

The intuition behind these definitions is the following.
A node $v_1 \in V'$ represents a partially defined configuration of $X_K$ where we have specified the contents of $V \setminus B^+$.
An edge to $v_2$ with label $S$ represents the action of shifting the configuration ``to the left'' by $\vec e_1$ and specifying the contents of the infinite strip $B^+ \cap B^-$ in the shifted version.
Since the configuration is $K$-periodic, it is enough to specify the contents of $D \cdot R$, which is exactly the domain of $S$.
Thus, a two-way infinite walk in the graph $G'$ determines a full $K$-periodic configuration.
Since $X$ is an SFT defined by $F$, it suffices to keep track of translates of $P \in F$ whose domain intersects $B^+ \cap B^-$.
The patterns are shifted to the left with the entire configuration.
They are not allowed to leave $B^+$ before being ``handled'' by disagreeing with the configuration on some element of $B^+ \cap B^-$, since then they would occur in the completely specified part of the configuration.

All in all, we have computed a finite digraph $G' = (V', E')$ whose edges are labeled by patterns in $A^{D \cdot R}$.
If we replace each label $S$ by its density $\sum_{v \in D(S)} W(v) / |D(S)| \in \R$, then the minimum weight $W(X_K)$ equals the minimum weight of a bi-infinite walk on $G'$.
The latter, in turn, is achieved by a simple cycle \cite[Lemma~4.1]{SaTo23}, so it suffices to compute the minimum density of a simple cycle in $G'$.
For that, we use Karp's minimum mean cycle algorithm \cite{Ka78}, which is readily parallelizable.

We perform some additional optimizations to reduce the size of the graph $G'$ labeled by $\R$ before applying Karp's algorithm.
Namely, we iterate the following operations until the graph no longer changes:
\begin{enumerate}
\item
Simplify the graph using a modified version of Hopcroft's algorithm for DFA minimization \cite{Ho71}, resulting in a potentially smaller graph that has the same set of labels of finite (and thus bi-infinite) walks.\footnote{Note that our graph is not \emph{right resolving} in the sense of \cite[Definition~3.3.1]{LiMa95}, which corresponds to being a DFA instead of an NFA. Thus the result may not be minimal in the sense of having the absolute smallest number of vertices of any graph with the same set of labels of walks. Nevertheless, the algorithm never increases the number of vertices, and in practice can substantially decrease it.}
\item
For each pair of vertices $v, v' \in V'$, remove all edges from $v$ to $v'$ except the one with the smallest weight.
\end{enumerate}
Minimization does not change the set of labels of bi-infinite walks on a graph, and the second operation does not change the minimum density of such walks, so they are safe for our purposes.

We have already used this functionality to find a new identifying code on the infinite hexagonal grid with density $53/126$ \cite{SaTo23}, down from the previous record of $3/7$ \cite{CoHoLoZe00}.
To reproduce the code, define the hexagonal grid and the SFT of identifying codes as in Section~\ref{sec:defining}, and run the commands
\begin{verbatim}
%compute_forbidden_patterns idcode 3
%minimum_density idcode threads=3 (5,1)
\end{verbatim}
Here, \texttt{threads=3} is an \emph{optional argument} specifying the number of threads (several commands accept some optional aruments, but \texttt{\%minimum\_density} is currently the only one to use parallel computation).
The computation takes about 17 minutes on a laptop computer.

\subsection{Lower bounds by discharging}

We then describe the algorithm for computing lower bounds for $W(X)$.
This algorithm does not depend on the existence of periodic points.
Given enough computational resources, it will give arbitrarily good approximations to $W(X)$.

The theoretical background is the \emph{discharging argument} used widely in graph theory \cite{CrWe17}.
We describe it in our context.
The idea is to re-distribute the weights of the nodes using deterministic local rules, and prove by a finitary argument that the resulting configuration of weights is always at least some constant $\alpha$ everywhere, implying $W(X) \geq \alpha$.
The \emph{initial charge} of a configuration $x \in A^V$ is the function $C_x : \Z^d \to \R$ defined as $C_x(\vec n) = \sum_{r \in R} W(x_{\vec n \cdot r}) / |R|$.
Given a topology $G = (V, E)$ with representative set $R \subset V$ and an alphabet $A$, a \emph{discharging rule} is a triple $(P, \vec m, c)$ where $P \in A^D$ is a finite pattern, $\vec m \in \Z^d$ and $c \in \R$.
A finite set $\mathcal{D}$ of discharging rules defines a new charge $\mathcal{D}(C_x)$ by
\begin{equation}
  \label{eq:discharging}
  \mathcal{D}(C_x)(\vec n) = C_x(\vec n) + \text{\raisebox{-0.5mm}{$\left( \text{\raisebox{2mm}{$\sum_{\substack{(P, \vec m, c) \in \mathcal{D} \\ ((\vec n - \vec m) \cdot x)|_{D(P)} = P}} c$}} \right)$}} - \text{\raisebox{-0.5mm}{$\left( \text{\raisebox{2mm}{$\sum_{\substack{(P, \vec m, c) \in \mathcal{D} \\ (\vec n \cdot x)|_{D(P)} = P}} c$}} \right)$}}.
\end{equation}
If $X \subset A^V$ is a $G$-subshift and $\mathcal{D}(C_x)(\vec n) \geq \alpha$ for all $x \in X$ and $\vec n \in \Z^d$, then $W(X) \geq \alpha$.
The prove that $\mathcal{D}(C_x)(\vec n) \geq \alpha$ always holds, it suffices to consider every pattern $Q \in A^N$ that occurs in $X$, where $N \subset V$ is finite but large enough to contain the domain of $P$ and $-\vec m \cdot P$ for all $(P, \vec m, c) \in \mathcal{D}$, and show that each of them satisfies $\mathcal{D}(C_x)(\vec 0) \geq \alpha$ whenever $x|_N = Q$.
This is a finite computation.

We use the idea of Stolee \cite{St14}, and employ a linear program to find a good set of discharging rules automatically. 
First, fix a finite domain $D \subset V$ and a finite set of vectors $T \subset \Z^d$.
Generate a set of patterns $\mathcal{P} \subset A^D$ that includes all patterns of shape $D$ that occur in $X$, by enumerating all locally allowed patterns over some larger finite domain.
Our triples will be $(P, \vec m, c(P, \vec m))$ for all $P \in \mathcal{P}$ and $\vec m \in T$, with $c(P, \vec m) \in \R$ being unconstrained variables in our linear program.
We also add another variable $\alpha \in \R$.

Next, generate another set of patterns $\mathcal{Q} \subset A^{D'}$ over the domain $D' = \bigcup_{\vec m \in T \cup \{\vec 0\}} -\vec m \cdot D$ that contains all $D'$-shaped patterns occurring in $X$.
Each pattern $Q \in \mathcal{Q}$ defines a constraint for the linear program, similarly to \eqref{eq:discharging}: the value of $\alpha$ must be less than or equal to the sum of $C_Q(\vec 0)$, each $c(P, \vec m)$ for which $(-\vec m \cdot Q)|_D = P$, and each $-c(P, \vec m)$ for which $Q|_D = P$.
Any set of values for $\alpha$ and the $c(P, \vec m)$ that satisfy the constraints gives a valid lower bound for $W(X)$.
Hence, we should maximize the value of $\alpha$ under the constraints.

A solution of the linear program is a best possible discharging strategy that transfers charge along the vectors $T$ based on occurrences of the patterns $\mathcal{P}$, given that the SFT $X$ may contain any of the patterns $\mathcal{Q}$.
Adding more nodes to the domain $D$ or more vectors to $T$, or refining the sets $\mathcal{P}$ or $\mathcal{Q}$, generally results in a better bound, but the computational cost will increase accordingly.
The associated Diddy command is \texttt{\%density\_lower\_bound radius domain vectors}.
As an example, we can compute a lower bound for the density of an identifying code on the hexagonal grid by \texttt{\%density\_lower\_bound idcode 0 (0,0,0) (0,0,1) (-1,0,1) (0,1,1) (0,-1,0) (1,0,0); (0,-1) (0,1) (1,0)}.
In about 65 seconds, Diddy produces the bound 0.4, same as in the article \cite{KaChLe98} that introduced the problem.
It has since been greatly improved, most recently to 23/55 \cite{St14}.

\section{Visualization}

Diddy also allows one to interactively generate and display locally valid finite patches of SFTs on $\Z^2$-like graphs.
The command \texttt{\%tiler name} opens a new window containing a finite grid-shaped subgraph of the topology $G$.
The user can zoom and pan the camera, set the values of any nodes, and ask Diddy to assign values to the remaining nodes so that the patch is locally valid in the SFT \texttt{name}, if possible.
We use the Pygame library \cite{pygame} for the visualization.

\section{Future directions}

Below is a partial list of features that the authors plan to implement.
\begin{itemize}
\item
  Support for general block maps between distinct SFTs.
\item
  Condition 2 of the SFT comparison algorithm could be extended to e.g.\ eventually periodic, semilinear or automatic configurations.
\item
Better support for the interplay between block maps, patterns and SFTs.
For example, one might want to search for invariant SFTs, temporally periodic points or attractors of a CA using a SAT solver.
\item
The syntax could be expanded to make various SFTs and CAs easier to implement.
These include distance calculations in graphs (for e.g.\ identifying codes of arbitrary radii), multi-layer SFTs and cellular automata (e.g.\ the Robinson tile set \cite{Ro71}), and definitions that involve counting or arithmetic (e.g.\ linear cellular automata over $\Z_p$, or Conway's Life).
\item
Many problems that are undecidable for two- and higher-dimensional SFTs can be solved effectively in the one-dimensional case, using finite automata theory and linear algebra. See e.g.~\cite[Sections~3.4~and~4.3]{LiMa95}.
Diddy should include total algorithms for this special case.
\item
The \emph{trace} of a 2D SFT is set of infinite columns occurring in its configurations.
It can be seen as a 1D subshift that can be approximated from the above (sometimes exactly) by sofic shifts.
Diddy should include functionality for extracting and analyzing traces of multidimensional SFTs.
\end{itemize}

\subsubsection*{Acknowledgements}
Ilkka Törmä was supported by the Academy of Finland under grant 346566.

%
%
%
\bibliographystyle{plainurl}
\bibliography{diddybib}

\begin{thebibliography}{10}

\bibitem{glucose}
Gilles Audemard and Laurent Simon.
\newblock {Glucose 4.1}, 2016.
\newblock URL: \url{https://www.labri.fr/perso/lsimon/research/glucose/}.

\bibitem{Be66}
Robert Berger.
\newblock The undecidability of the domino problem.
\newblock {\em Mem. Amer. Math. Soc. No.}, 66, 1966.
\newblock 72 pages.

\bibitem{CoHoLoZe00}
G\'{e}rard~D. Cohen, Iiro Honkala, Antoine Lobstein, and Gilles Z\'{e}mor.
\newblock Bounds for codes identifying vertices in the hexagonal grid.
\newblock {\em SIAM J. Discrete Math.}, 13(4):492--504, 2000.
\newblock \href {https://doi.org/10.1137/S0895480199360990}
  {\path{doi:10.1137/S0895480199360990}}.

\bibitem{CrWe17}
Daniel~W. Cranston and Douglas~B. West.
\newblock An introduction to the discharging method via graph coloring.
\newblock {\em Discrete Math.}, 340(4):766--793, 2017.
\newblock \href {https://doi.org/10.1016/j.disc.2016.11.022}
  {\path{doi:10.1016/j.disc.2016.11.022}}.

\bibitem{Ho71}
John Hopcroft.
\newblock An {$n$} log {$n$} algorithm for minimizing states in a finite
  automaton.
\newblock In {\em Theory of machines and computations ({P}roc. {I}nternat.
  {S}ympos., {T}echnion, {H}aifa, 1971)}, pages 189--196. Academic Press, New
  York, 1971.

\bibitem{pysat}
Alexey Ignatiev, Antonio Morgado, and Joao Marques{-}Silva.
\newblock {PySAT:} {A} {Python} toolkit for prototyping with {SAT} oracles.
\newblock In {\em SAT}, pages 428--437, 2018.
\newblock \href {https://doi.org/10.1007/978-3-319-94144-8\_26}
  {\path{doi:10.1007/978-3-319-94144-8\_26}}.

\bibitem{Ka78}
Richard~M. Karp.
\newblock A characterization of the minimum cycle mean in a digraph.
\newblock {\em Discrete Mathematics}, 23(3):309--311, 1978.
\newblock \href {https://doi.org/https://doi.org/10.1016/0012-365X(78)90011-0}
  {\path{doi:https://doi.org/10.1016/0012-365X(78)90011-0}}.

\bibitem{KaChLe98}
Mark~G. Karpovsky, Krishnendu Chakrabarty, and Lev~B. Levitin.
\newblock On a new class of codes for identifying vertices in graphs.
\newblock {\em IEEE Trans. Inform. Theory}, 44(2):599--611, 1998.
\newblock \href {https://doi.org/10.1109/18.661507}
  {\path{doi:10.1109/18.661507}}.

\bibitem{LiMa95}
Douglas Lind and Brian Marcus.
\newblock {\em An introduction to symbolic dynamics and coding}.
\newblock Cambridge University Press, Cambridge, 1995.
\newblock \href {https://doi.org/10.1017/CBO9780511626302}
  {\path{doi:10.1017/CBO9780511626302}}.

\bibitem{pygame}
{Pygame development team}.
\newblock {Pygame 2.3.0}, 2023.
\newblock URL: \url{https://github.com/pygame}.

\bibitem{Ro71}
Raphael~M. Robinson.
\newblock Undecidability and nonperiodicity for tilings of the plane.
\newblock {\em Invent. Math.}, 12:177--209, 1971.

\bibitem{pulp}
J.S. Roy, Stuart~A. Mitchell, Christophe-Marie Duquesne, and Franco Peschiera.
\newblock {PuLP 2.7.0}, 2022.
\newblock URL: \url{https://github.com/coin-or/pulp}.

\bibitem{SaTo22uniqf}
Ville Salo and Ilkka T\"{o}rm\"{a}.
\newblock What can oracles teach us about the ultimate fate of life?
\newblock In {\em 49th {EATCS} {I}nternational {C}onference on {A}utomata,
  {L}anguages, and {P}rogramming}, volume 229 of {\em LIPIcs. Leibniz Int.
  Proc. Inform.}, pages Art. No. 131, 20. Schloss Dagstuhl. Leibniz-Zent.
  Inform., Wadern, 2022.
\newblock \href {https://doi.org/10.4230/lipics.icalp.2022.131}
  {\path{doi:10.4230/lipics.icalp.2022.131}}.

\bibitem{SaTo22goe}
Ville Salo and Ilkka T\"{o}rm\"{a}.
\newblock Gardens of {E}den in the {G}ame of {L}ife.
\newblock In {\em Automata and complexity---essays presented to {E}ric {G}oles
  on the occasion of his 70th birthday}, volume~42 of {\em Emerg. Complex.
  Comput.}, pages 399--415. Springer, Cham, [2022] \copyright 2022.
\newblock \href {https://doi.org/10.1007/978-3-030-92551-2\_22}
  {\path{doi:10.1007/978-3-030-92551-2\_22}}.

\bibitem{diddyrepo}
Ville Salo and Ilkka Törmä.
\newblock {Diddy}, 2023.
\newblock URL: \url{https://github.com/ilkka-torma/diddy}.

\bibitem{SaTo23}
Ville Salo and Ilkka Törmä.
\newblock Finding codes on infinite grids automatically, 2023.
\newblock \href {http://arxiv.org/abs/2303.00557} {\path{arXiv:2303.00557}}.

\bibitem{St14}
Derrick Stolee.
\newblock Automated discharging arguments for density problems in grids, 2014.
\newblock \href {http://arxiv.org/abs/1409.5922} {\path{arXiv:1409.5922}}.

\bibitem{Wa61}
Hao Wang.
\newblock Proving theorems by pattern recognition {II}.
\newblock {\em Bell System Technical Journal}, 40:1--42, 1961.

\end{thebibliography}
\end{document}